\documentclass[12pt]{article}
\usepackage{epsfig}

\def\beq{\begin{equation}}
\def\eeq#1{\label{#1}\end{equation}}
\def\eeqn{\end{equation}}

\def\beqa{\begin{eqnarray}}
\def\eeqa#1{\label{#1}\end{eqnarray}}
\def\eeqan{\end{eqnarray}}

\let\bar=\overbar

\def\Dslash{\not{\hbox{\kern-4pt $D$}}}
\def\dslash{\not{\hbox{\kern-2pt $\del$}}}

\def\msb{{\bar{\ssstyle M \kern -1pt S}}}

\def\kpipi{$D^+\rightarrow K^-\pi^+\pi^+$\ }
\def\kpp{$K^-\pi^+\pi^+$\ }
\def\ka14{$K^*_0(1430)$\ }
\def\d3pi{$D^+\rightarrow \pi^-\pi^+\pi^+$\ }
\def\ds3pi{$D_s^+\rightarrow \pi^-\pi^+\pi^+$\ }
\def\Title#1{\begin{center} {\Large {\bf #1} } \end{center}}

\begin{document}

\Title{Light  Scalar Mesons  in Charm Meson Decays }

\bigskip\bigskip

\begin{raggedright}  

{{\it Ignacio Bediaga} \\
 Representing the Fermilab E791 Collaboration \\
Centro Brasileiro de Pesquisas F\'\i sicas, \\
Rua Xavier Sigaud 150, 22290, Rio de Janeiro, Brazil \\
bediaga@cbpf.br}

\bigskip\bigskip

\end{raggedright}

\begin{abstract}
We present recent results on scalar light mesons based on Dalitz plot 
analyses of charm decays from Fermilab experiment E791. Low mass scalar mesons 
are found to have large contributions to the decays studied, $D^+\to K^-\pi^+\pi^+$ and 
$D^+, D_s^+\to\pi^-\pi^+\pi^+$. These results demonstrate the importance  
of charm decays as a new environment for the study of light meson physics.

\end{abstract}

\section{Introduction}

Here we present an overview of the  results we obtained  analysing  
the decays \ds3pi 
\cite{ds3pi}, \d3pi \cite{d3pi} and  \kpipi \cite{kpipi}, using data 
collected in 1991/92 by Fermilab experiment E791 from $500
  GeV/c \pi- -nucleon$ interactions.  For details see [ref791].

To obtain these results, we had to introduce a new approach in Dalitz plot 
analysis in order to extract the mass and width of the scalar resonances 
by allowing them as floating parameters in the fit.  
We begin this paper presenting the general  method, applied in the \kpipi 
Dalitz-plot analysis, then we discuss the \ds3pi and \d3pi studies using the 
same procedure.

\section{The \kpipi Dalitz-plot Analysis}

\begin{figure}[t]
\centerline{
\begin{minipage}{2.5in}
\epsfxsize=12pc
\epsfbox{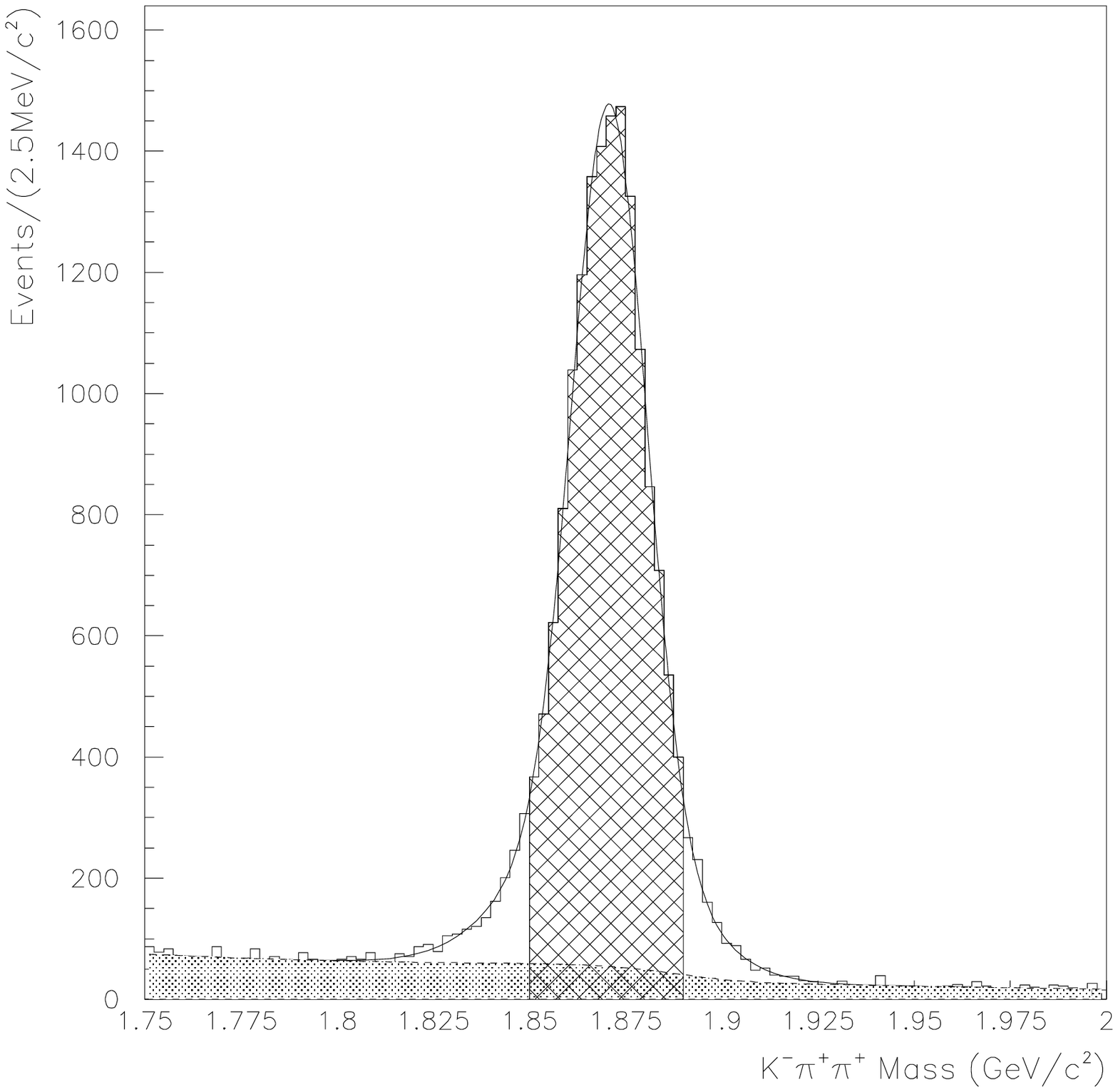}
\end{minipage}
\begin{minipage}{2.5in}
\epsfxsize=12pc
\epsfbox{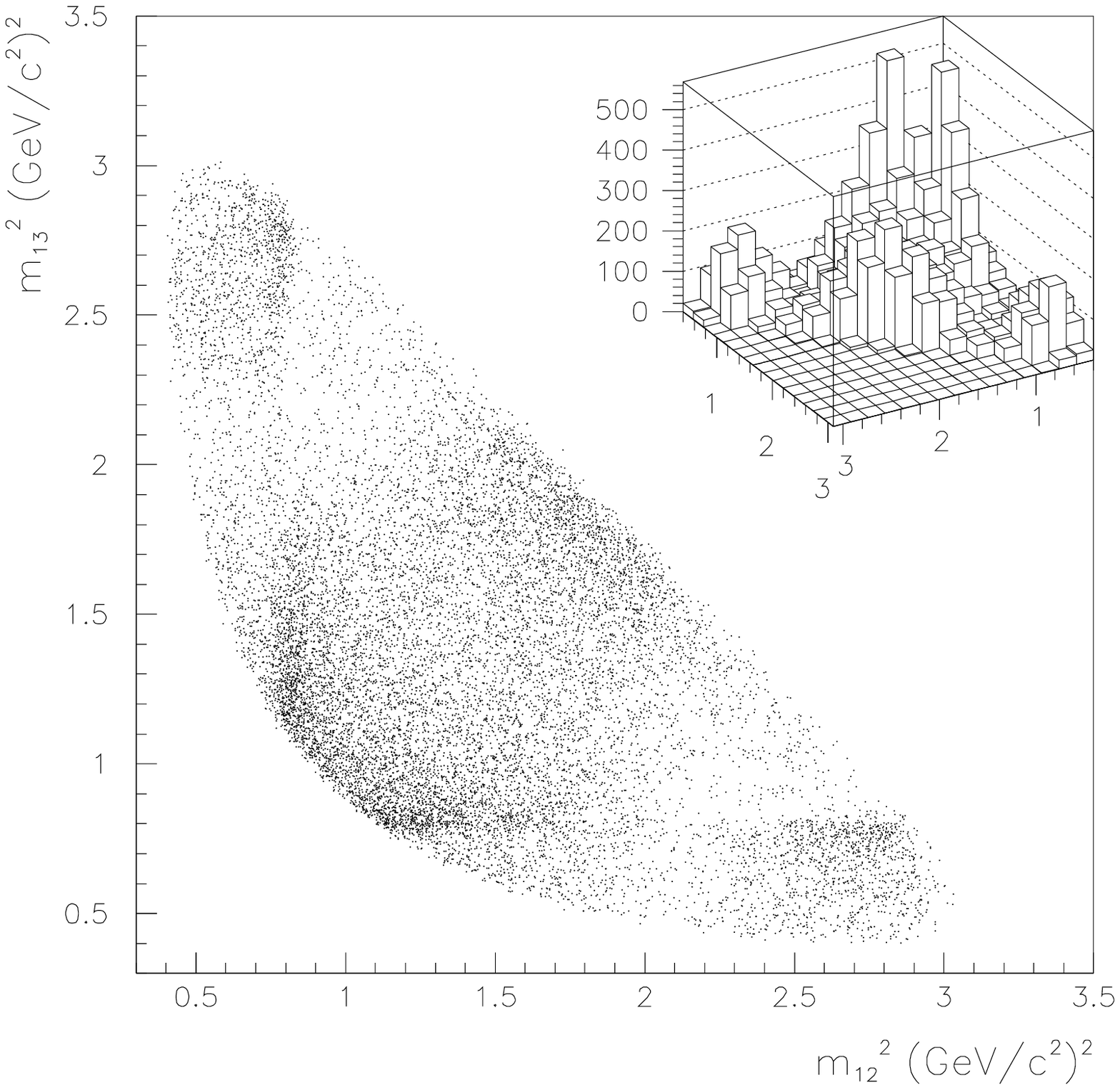}
\end{minipage}}
\caption{(a) The \kpp invariant mass spectrum. The filled area is 
background; (b) Dalitz plot corresponding to the events in the dashed area of (a).
\label{kpipi} }
\end{figure}

From the  original $2\times 10^10$ events collected by E791, and 
after reconstruction and selection criteria, we obtained the $k^-\pi^+\pi^+$
 sample shown in Figure~\ref{kpipi}(a).  The cross-hatched region contains the events selected 
for the Dalitz-plot ana\-lysis. There are 15090 events in this sample, of which 
6\% are background. 

Figure~\ref{kpipi}(b) shows the Dalitz-plot for these events. The plot presents 
a rich structure, where we can observe the clear
bands from $\bar K^*(890)\pi^+$, and an accumulation of events at the upper edge of
the diagonal, due to heavier resonances.
To study the resonant substructure, we perform an unbinned maximun-likelihood fit to the 
data, with probability distribution functions (PDF's) for both signal and background sources.
In particular, for each candidate event, the signal PDF is written as the square of the 
total physical amplitude ${\cal A}$  and it is weighted
for the acceptance across the Dalitz plot (obtained by Monte Carlo (MC)) and by 
the level of signal to background for each event, as given by the line shape of 
Figure~\ref{kpipi}(a). The background PDF's (levels and shapes) are fixed for the Dalitz-plot fit, 
according to MC and data studies.

We begin describing our first approach to fit the data, which represents the conventional
Dalitz-plot analysis including the known $K\pi$ resonant amplitudes (${\cal A}_n,~n\ge 1$), plus 
a constant non-resonant contribution. The resonance  amplitudes used to 
describe the signal, Breit-Wigner parametrizations with  Blatt-Weisskopf 
damping  factors are  described in reference \cite{kpipi}.

Using this model with well-known resonances, we find contributions from the following channels: the non-resonant,
responsible for more than 90\% of the decay rate, followed by $\bar K^*_0(1430)\pi^+$, 
$\bar K^*(892)\pi^+$, $\bar K^*(1680)\pi^+$ and $\bar K^*_2(1430)\pi^+$. The decay fractions 
and relative phases are available in reference \cite{kpipi}.

To evaluate the fit quality, we compute a $\chi^2$ from binned, 
two-dimensional distributions of data and decay model events. 
The $\chi^2$ comes from the differences in the binned numbers of events between
the data and the model (from a fast MC simulation).
We obtain $\chi^2/\nu=2.7$ ($\nu$ being the number of degrees of freedom), with a corresponding 
confidence level (CL) of $10^{-11}$. We thus conclude that a model with 
the known $K\pi$ resonances, plus a non-resonant amplitude, is not able to describe 
the \kpipi Dalitz plot satisfactorily. Thus, we are led to try an extra scalar 
resonance in our fit model. This second fit model, is constructed 
by the inclusion of an extra scalar state with unconstrained mass
and width. For consistency, the mass and width of the other scalar state, 
the $K^*_0(1430)$, are also free parameters of the fit. We adopt a better 
description for these scalar states by introducing gaussian-type form-factors 
\cite{torn2} to take into account the finite size of the decaying mesons. 

Using this model, we obtain the values of  $797\pm 19\pm 42$ MeV/c$^2$ for the mass and 
$410\pm 43\pm 85$ MeV/c$^2$ for the width of the new scalar state (first error 
statistical, second error systematic), referred to here as the $\kappa$.
The values of mass and width obtained for the $K^*_0(1430)$ are respectively 
$1459\pm 7\pm 6$ MeV/c$^2$ and $175\pm 12\pm 12$ MeV/c$^2$, appearing heavier and
narrower than presented by the PDG \cite{pdg}. The decay fractions and relative phases 
 with systematic errors, are given in reference \cite{kpipi}. 
 The $\kappa\pi^+$ state is now the dominant channel with decay fraction 
about 50\%.

\section{The \ds3pi Results}

\begin{figure}[t]
\epsfxsize=18pc
\centerline{
\epsfbox{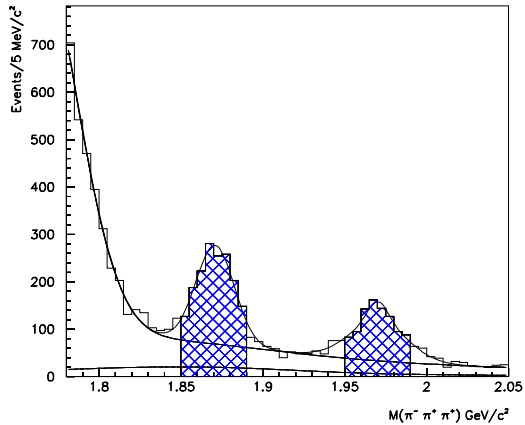}} 
\caption{The $\pi^-\pi^+\pi^+$ invariant mass spectrum. The dashed line represent  
the total background. Events used for the Dalitz analyses
are in the hatched areas.}
\label{m3pi} 
\end{figure}

\begin{figure}[t]
\centerline{
\begin{minipage}{2.5in}
\epsfxsize=10pc
\epsfbox{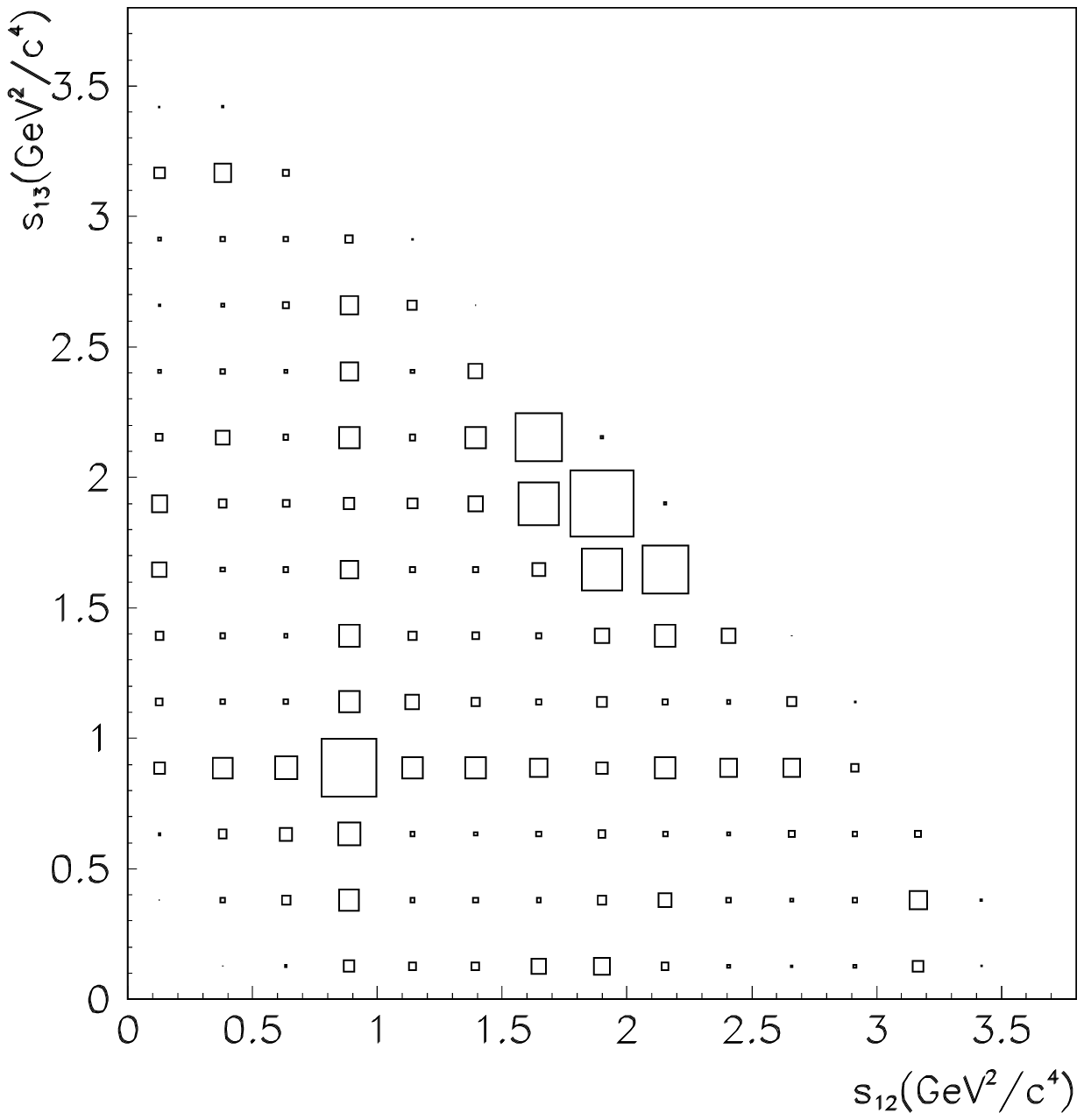}
\end{minipage}
\begin{minipage}{2.5in}
\epsfxsize=10pc
\epsfbox{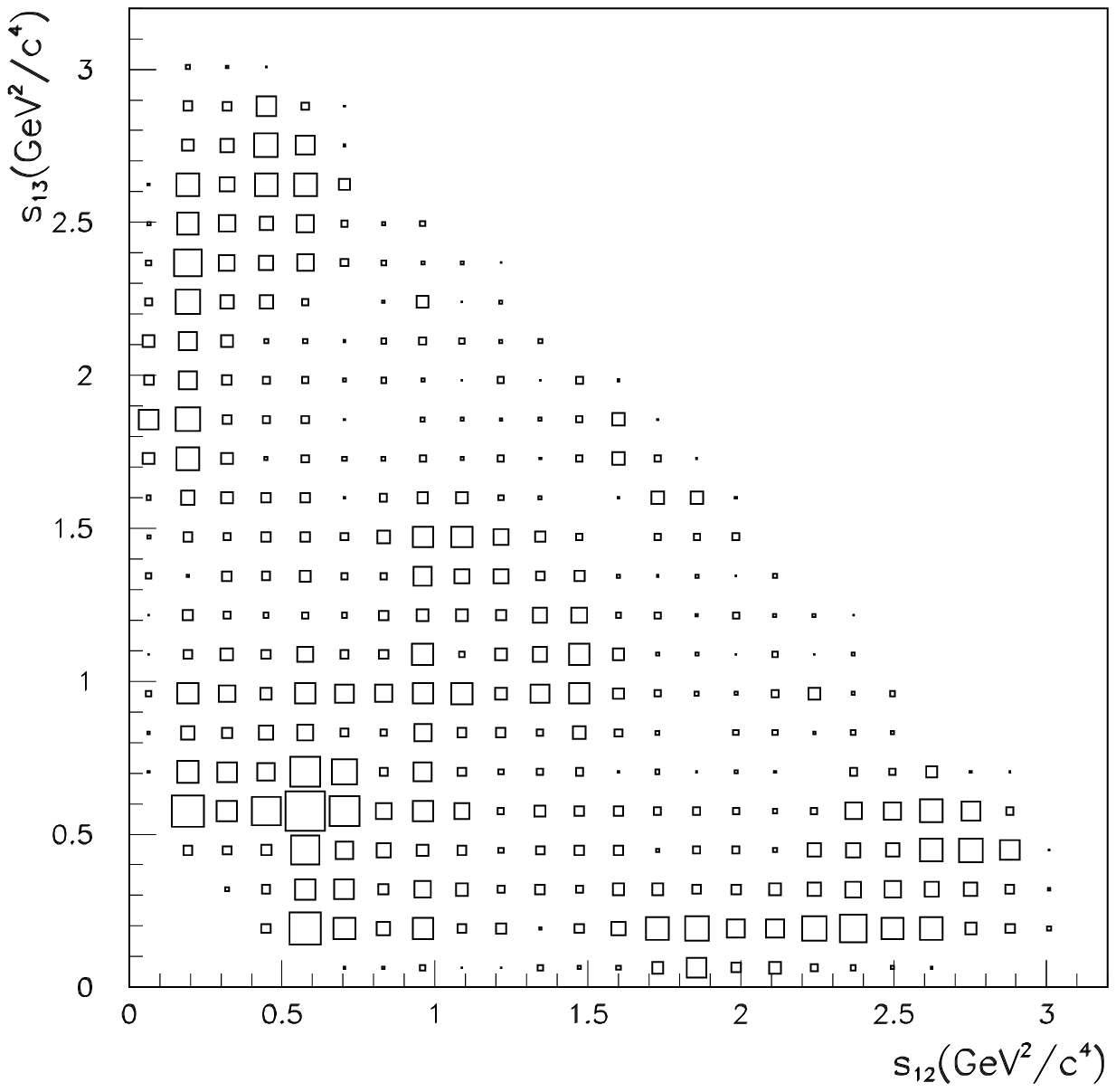}
\end{minipage}}
\caption{(a) The $D_s^+ \to \pi^- \pi^+ \pi^+$ Dalitz plot and
(b) the \d3pi Dalitz plot. Since there are two identical pions, 
the plots are symmetrized.
\label{dalitz3pi} } 
\end{figure}

In Figure~\ref{m3pi} we show the $\pi^-\pi^+\pi^+$ invariant mass distribution 
for the sample collected by E791 after reconstruction and selection 
criteria \cite{ds3pi,d3pi}. Besides combinatorial background, reflections 
from the decays \kpipi, $D^0\to K^-\pi^+$ 
(plus one extra $\pi^+$ track) and $D_s^+\to \eta'\pi^+, ~\eta'\to\rho^0(770)\gamma$ 
are  all taken into account. The hatched regions in Figure \ref{m3pi} show 
the samples used for the Dalitz-plot analyses. There are 1686 and 937 
candidate events for $D^+$  and $D_s^+$ respectively, with a signal to 
background ratio of about 2:1.  The Dalitz plots for these events are shown 
in Figure~\ref{dalitz3pi}, the axes corresponding to the two $\pi^-\pi^+$ 
invariant-masses 
squared.

For the \ds3pi events in Figure~\ref{dalitz3pi}(a), the signal amplitude includes
all channels with well known dipion resonances \cite{pdg}: $\rho^0(770) \pi^+$, 
$f_0(980) \pi^+$, $f_2(1270) \pi^+$, $f_0(1370) \pi^+$, $\rho^0(1450) \pi^+$ 
and the non-resonant, assumed constant across the Dalitz plot.

The measured $f_0(980)$ standard Breit-Wigner  parameters are 
 $m_0 = 975 \pm 3$ MeV/c$^2$ and $\Gamma_0 =  44 \pm 2 \pm 2$ MeV/c$^2$.
The confidence level of the fit for \ds3pi is 35\%.The decay fractions 
and relative phases, with systematic errors, are given in reference \cite{ds3pi}. 
 The $f_0(980)\pi^+$ state is the dominant channel with decay fraction 
 about 50\%. 
 
\begin{figure}[htb]
\centerline{
\begin{minipage}{3.in}
\epsfxsize=15pc
\epsfbox{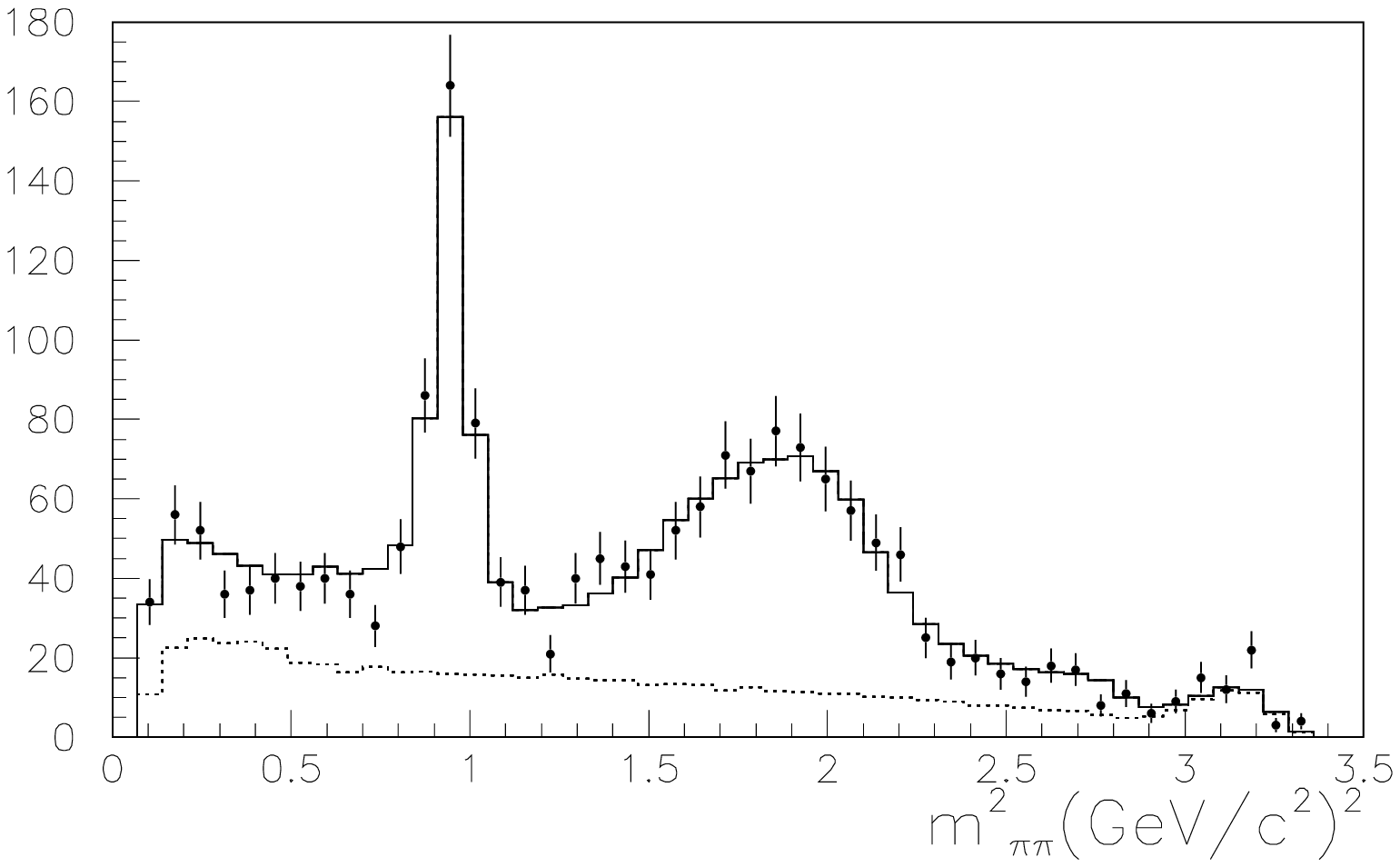}
\end{minipage}
\begin{minipage}{2.5in}
\epsfxsize=15pc
\epsfbox{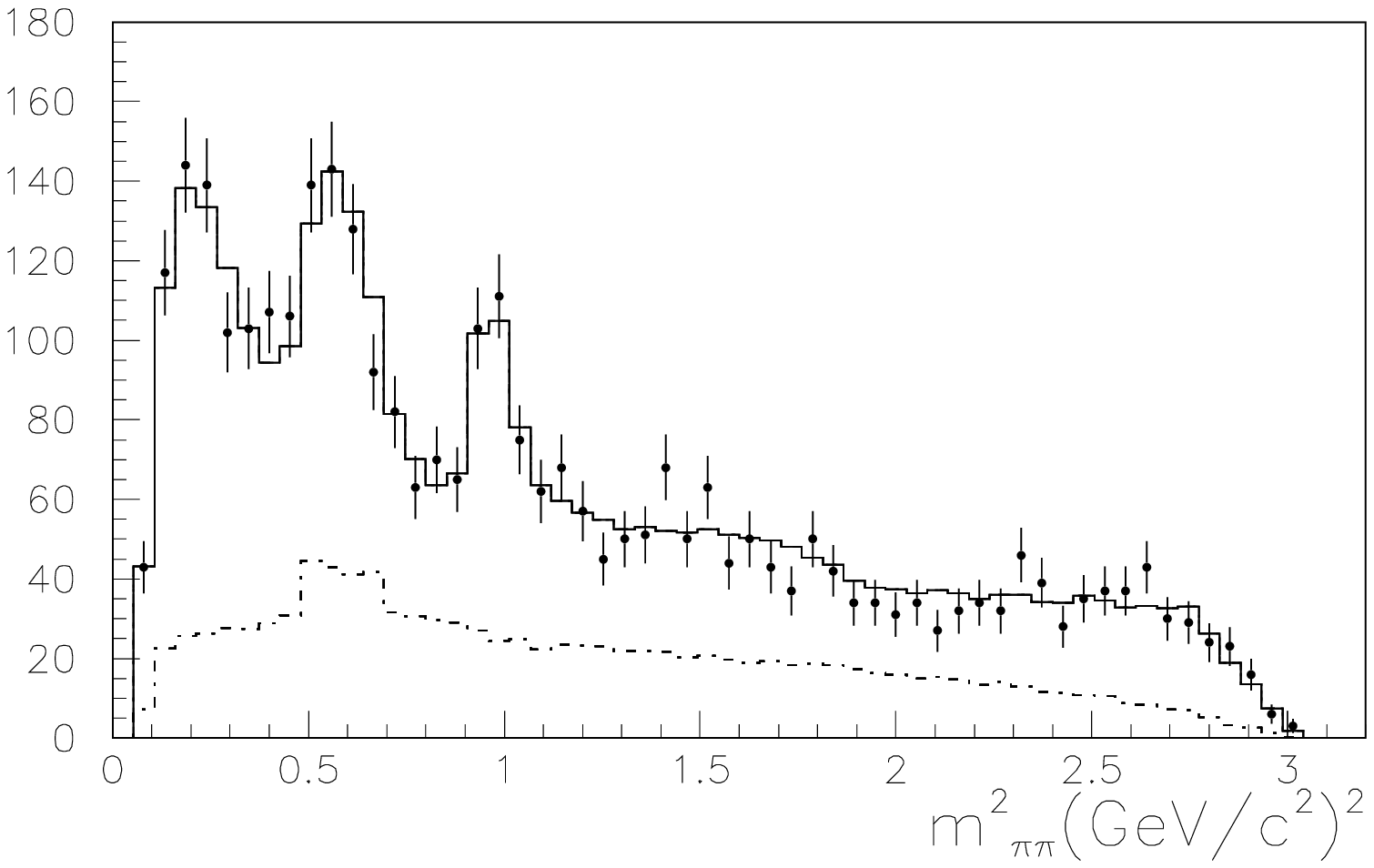}
\end{minipage}}
\caption{ (a) $s_{12}$ and $s_{13}$ ($m^2_{\pi\pi}$) projections for 
\ds3pi  data (dots) and our best fit
(solid). (b) $s_{12}$ and $s_{13}$  projections for \d3pi data (dots) 
and our best fit (solid) for models  with $\sigma(500)\pi^+$ amplitude. 
In both figures the dashed distribution corresponds to the expected background level.}
\label{proj_d3pi} 
\end{figure}

In Figure~\ref{proj_d3pi} we show
the $\pi^-\pi^+$ mass-squared projections for data (points)
and model (solid lines, from fast-MC).

\section{The \d3pi Results}
\label{secd3pi}

In a first approach, we try to fit the \d3pi Dalitz plot of Figure~\ref{dalitz3pi}(b)
with the same amplitudes used for the \ds3pi analysis. Using this model, 
the non-resonant, the $\rho^0(1450)\pi^+$, and the $\rho^0(770)\pi^+$ amplitudes
are found to dominate \cite{d3pi}. However, this model does not describe the data 
satisfactorily, especially at low 
$\pi^-\pi^+$ mass squared \cite{d3pi}. The $\chi^2/\nu$  obtained from the binned Dalitz plot 
for this model is 1.6, with a CL less than $10^{-5}$. 
   
To investigate the possibility that another $\pi^-\pi^+$ resonance contributes to the
\d3pi decay, we add an extra scalar resonance amplitude to the signal PDF, with  
mass and width as floating parameters in the fit.

We find that this model improves our fit substantially. 
The mass and the width of the extra scalar state are found to be 
$ 478^{+24}_{-23} \pm 17  $ MeV/$c^2$ and $ 324^{+42}_{-40}  \pm 21$ MeV/$c^2$, 
respectively. Refering to this state as the $\sigma(500)$, we observe that the
$\sigma(500)\pi^+$ channel produces the largest decay fraction \cite{d3pi}; 
the non-resonant amplitude, which is dominant in the model without 
$\sigma(500)\pi^+$, drops substantially. 
The model with the sigma  describes the data very well, as can be seen by the $\pi\pi$ mass squared
projection in Fig.~\ref{proj_d3pi}(b). The $\chi^2/\nu$ is now 0.9, with a 
corresponding  confidence level of 91\%.


\section{Conclusion}

From the data of the Fermilab E791 experiment, we studied the Dalitz plots of 
the decays $D^+\to K^-\pi^+\pi^+$, \ds3pi, and $D^+\rightarrow \pi^-\pi^+\pi^+$. 
In these three final states, the scalar intermediate resonances were found 
to give the largest contributions to the decay rates. We obtained strong 
evidence for the existence of the $\sigma(500)$ and $\kappa$ scalar mesons, 
measuring their masses and widths. We also obtained new measurements for masses and
widths of the other scalars studied, $f_0(980)$, $f_0(1430)$ and $K^*_0(1430)$.

The results presented here show the potential of $D$ meson decays for the study of 
light meson spectroscopy, in particular in the scalar sector \cite{torn3}.

\end{document}